\documentclass[preprint2]{aastex}
\usepackage[makeroom]{cancel}
\usepackage{graphicx}
\usepackage{epstopdf}
 \usepackage{amssymb, latexsym, amsmath}
 \usepackage[dvips]{color}
 \usepackage{graphicx}
  \usepackage{color}

 \newcommand{\be}{\begin{equation}}
 \newcommand{\ee}{\end{equation}}
 \newcommand{\ba}{\begin{eqnarray}}
 \newcommand{\ea}{\end{eqnarray}}
 \newcommand{\bs}{\begin{subequations}}
 \newcommand{\es}{\end{subequations}}

 \newcommand{\ubold}{{\bf{u}}}

  \newcommand{\vbold}{{\bf{v}}}

 \slugcomment{Submitted to the Astrophysical Journal}
 \shorttitle{Tidal heating}
 \shortauthors{Efroimsky \& Makarov}

\begin{document}
  \title{${{~~~~~~~~~~~~~~~~~~~~~~~~~~~~~~~~~~}^{
          {
            {
            Published~in~
                the~Astrophysical~Journal
        \,,~Vol.~795\,,~id.~7\,~(2014)
                  }}}}$
                  ~\\
 Tidal dissipation in a homogeneous spherical body.\\
 $ \, $II.$ \, $ Three examples: Mercury, $ \, $~Io, $ \, $ and Kepler-10$ \, $b}
\author{Valeri V. Makarov $ \, $and$ \, $ Michael Efroimsky ~~~}
\affil{US Naval Observatory, 3450 Massachusetts Avenue NW, Washington DC 20392}
\email{vvm@usno.navy.mil$ \, $, ~~michael.efroimsky@usno.navy.mil}
     \date{}

 \begin{abstract}
 In our recent study (Efroimsky \& Makarov 2014),
 we derived from the first principles a formula for the tidal heating rate in a tidally perturbed homogeneous sphere. We compared our result with the formulae used in the literature, and we pointed out the differences. Now, using this result, we present three case studies  -- Mercury, the enigmatic Kepler-10$ \, $b, and a triaxial Io. A very sharp frequency dependence of $\,k_2/Q\,$ near spin-orbit  resonances yields a similarly sharp dependence of $\,k_2/Q\,$ (and, therefore, of tidal heating) upon the spin rate. This indicates that physical libration may play a major role in the tidal heating of synchronously rotating planets. The magnitude of libration in the spin rate being defined by the planet's triaxiality, the latter should be a factor determining the dissipation rate. Other parameters equal, a synchronously rotating body with a stronger triaxiality should generate more heat than a similar body of a more symmetrical shape. Further in the paper, we discuss possible scenarios where initially triaxial objects melt and lose their triaxiality. Thereafter, dissipation in them becomes less intensive; so the bodies solidify. The tidal bulge becomes a new permanent figure, with a new triaxiality lower than the original. In the paper, we also derive simplified, approximate expressions for the dissipation rate in a rocky planet of the Maxwell rheology, with a not too small Maxwell time (longer than the inverse tidal frequency). The three expressions derived pertain to the cases of a synchronous spin, a 3:2 resonance, and a nonresonant rotation; so they can be applied to most close-in super-Earth exoplanets detected thus far. In such bodies, the rate of tidal heating outside of synchronous rotation is weakly dependent on the orbital eccentricity and equator's obliquity, provided both these parameters are small or moderate. According to our calculation, the rocky Kepler-10$ \, $b, which is one of the densest exoplanets known to date, could hardly survive the great amount of tidal heating without being synchronised, circularised and also reshaped through a complete or partial melt-down.\\
 ~\\
 \end{abstract}

 \pagebreak

 \section{Motivation and plan}

 In the work by Efroimsky \& Makarov (2014),
 we derived from first principles a formula for the tidal dissipation rate in a homogeneous spherical body. When restricted to the special case of an incompressible body spinning synchronously, that result was compared to the commonly used expression from Peale \& Cassen (1978, Eqn. 31), and the differences were pointed out. Now, using the theoretical exposition from Efroimsky \& Makarov (2014), we demonstrate how tidal dissipation can be estimated for synchronous and asynchronous rocky planets.

 Section \ref{brief} serves to remind the said expression for the tidal dissipation rate. It is compared with the analogous formulae from Kaula (1964) and Peale \& Cassen (1978).

 Section \ref{limit.sec} gives an overview of the popular simplified formulae derived from the theory by Peale \& Cassen and explains the highly restrictive conditions, under which these formulae can be used.

 Section \ref{merc.sec} presents the first example, Mercury. We show that tidal heating is not likely to have played a major role in the history of this planet, despite its considerable eccentricity and the fact that Mercury is in the 3:2 spin-orbit resonance.

 Sections \ref{io.seq} addresses the second practical example, tidal heating in Io. We provide arguments in favour of a hypothesis that the energy damping rate in synchronous bodies may be sensitive to triaxiality. This sensitivity stems from a very sharp, kink-shaped frequency-dependence of $\,k_2/Q\,$ near resonances, which is within the range of physical libration for significant values of triaxiality. Our hypothesis bears a qualitative character and should be propped up by numerical modeling, which will be presented elsewhere.

 Section \ref{case3} is devoted to the third example, Kepler-10$\,$b, a very dense super-Earth that may sooner be classified as a super-Mercury (Selsis et al. 2013). Given the extreme proximity of the planet to its host star (less than 0.017 AU), we presume that the planet is experiencing a considerable tidal interaction and may, therefore, be overheated. The mantle's response in this case is viscoelastic and may be approximated with the Maxwell model. Assuming finite values of eccentricity and equator obliquity, we estimate the rate of energy dissipation in Kepler-10$ \, $b, for the case of synchronism and for
 other rotational states. Tidal heating in this planet becomes so intense that the temperature should be increasing by several degrees per year, if the eccentricity is pumped up by the companion planet. We complete this section by sketching possible scenarios of rotational and thermal evolution of such close-in planets subject to extreme tides, including episodic melt-down and reshaping of their surfaces.

 In Section \ref{eqs.sec},
 we provide three simplified, approximate expressions for the dissipation rate: one for a synchronised planet, another for a planet in a nonresonant rotation, and a third for a planet trapped in the 3:2 spin-orbit resonance. These formulae are derived for a specific case when the rheology is viscoelastic (Maxwell, with no Andrade creep) and the Maxwell time is not too small (larger than the inverse tidal frequency).

 \section{Tidal dissipation of energy}\label{brief}

 Consider a planet of mass $\,M\,$ that is tidally disturbed by an external body of mass $\,M^{\,*}\,$. As seen from the planet, the perturber describes an orbit
 parameterised by the Keplerian variables $\,a,\,e,\,i,\,\omega,\,\Omega,\,{\cal{M}}\,$, which are: the semimajor axis, eccentricity, inclination, agrument of the pericentre, longitude of the node, and mean anomaly.

 In the frame of the planet, the external tide-raising potential can be expanded in a Fourier series whose terms will contain sines and cosines of $\;\omega_{\textstyle{_{lmpq}}}\,t\;$. Here $\,t\,$ is time and $\,\omega_{\textstyle{_{lmpq}}}\,$ are the Fourier tidal modes. As explained, e.g., in Efroimsky \& Makarov (2013), these are given by
 \be
 \omega_{\textstyle{_{lmpq}}}~=\;(l-2p)\;\dot{\omega}\,+\,(l-2p+q)\;{\bf{\dot{\cal{M}}}}\,+\,m\;(\dot{\Omega}\,-\,\dot{\theta})~~,
 \label{moda}
 \ee
 $lmpq\,$ being integers, $\,\theta\,$ and $\,\dot{\theta}\,$ being the rotation angle and spin rate of the disturbed body, and ${\bf{\dot{\cal{M}}}}\,$ being the perturber's ``anomalistic" mean motion.
 While the Fourier modes $\,\omega_{\textstyle{_{lmpq}}}\,$ can assume either sign, the resulting physical forcing frequencies are positive definite:
 \begin{eqnarray}
 \chi_{\textstyle{_{lmpq}}} \,=\,|\,\omega_{\textstyle{_{lmpq}}}\,|
 \,~.
 \label{2}
 \end{eqnarray}
 In Efroimsky \& Makarov (2014), we derive a general formula for the time-averaged damping rate. When the apsidal precession of the perturber, as seen from the perturbed body, is uniform, the rate is:
  \ba
  \nonumber
  \langle\,P\,\rangle\,=\,
    \frac{G\,{M^*}^{\,2}}{a  }\sum_{l=2}^{\infty}\left(\frac{R\,}{a}\right)^{\textstyle{^{2l+1}}}\sum_{m=0}^{l}
    \frac{(l - m)!}{({\it l} + m)!}
  \left(2-\delta_{0m}\right)
  \nonumber
  \ea
  \be
  \sum_{p=0}^{l}F^{\,2}_{lmp}(i)
    \sum_{q\,=-\infty}^{\infty}G^{\,2}_{lpq}(e)
  \,\omega_{\textstyle{_{lmpq}}}\,
  k_l(\omega_{\textstyle{_{lmpq}}})~\sin\epsilon_l(\omega_{\textstyle{_{lmpq}}})~,\,~~
  \label{3}
  \ee
  where $\,k_l(\omega_{\textstyle{_{lmpq}}})\,$ and $\,\epsilon_l(\omega_{\textstyle{_{lmpq}}})\,$ are the dynamical Love numbers and tidal phase lags.

   Being even functions of the tidal modes, the dynamical Love numbers may as well be understood as functions of the physical frequencies (\ref{2}):
 \begin{eqnarray}
 k_l(\omega_{\textstyle{_{lmpq}}})~=~k_l(\chi_{\textstyle{_{lmpq}}})~~.
 \label{}
 \end{eqnarray}
 The phase lags are odd functions of $\,\omega_{\textstyle{_{lmpq}}}\,$ and have the same sign as $\,\omega_{\textstyle{_{lmpq}}}\,$. So they may be written down as
 \begin{eqnarray}
 \epsilon_l(\omega_{\textstyle{_{lmpq}}})&=&|\,\epsilon_l(\omega_{\textstyle{_{lmpq}}})\,|~\mbox{Sgn}\,\omega_{\textstyle{_{lmpq}}}
 \nonumber\\
 \nonumber\\
 &=&
 \epsilon_l(\chi_{\textstyle{_{lmpq}}})\,~\mbox{Sgn}\,\omega_{\textstyle{_{lmpq}}}~~,~~~~~~
 \label{}
 \end{eqnarray}
 where $\,\epsilon_l(\chi_{\textstyle{_{lmpq}}})\,$ are non-negative, because so are $\,\chi_{\textstyle{_{lmpq}}}\,$. All in all, we have:
 \begin{eqnarray}
 k_l(\omega_{\textstyle{_{lmpq}}})~\sin\epsilon_l(\omega_{\textstyle{_{lmpq}}})\qquad\qquad\qquad\qquad\qquad\nonumber\\
 \nonumber\\
 =~k_l(\chi_{\textstyle{_{lmpq}}})~\sin\epsilon_l(\chi_{\textstyle{_{lmpq}}})~\,\mbox{Sgn}\,\omega_{\textstyle{_{lmpq}}}~~,~~~
 \label{6a}
 \label{6}
 \end{eqnarray}
 where $\,k_l(\chi_{\textstyle{_{lmpq}}})~\sin\epsilon_l(\chi_{\textstyle{_{lmpq}}})\,$ are positive definite and are often denoted as $\,k_l/Q_l\,$.

 The frequency dependence $\,k_l(\chi_{\textstyle{_{lmpq}}})$ $\sin\epsilon_l(\chi_{\textstyle{_{lmpq}}})\,$ is derived in the Appendix. It is a functional of the planet's rheology and also of its size and mass. At lower frequencies, self-gravitation is playing a key role in tidal damping, so the tidal quality factor{\underline{{s}}} defined through $\,1/Q_l=\sin\epsilon_l(\chi)\,$ differ considerably from the seismic quality factor $\,Q\,$. However, they approach $\,Q\,$ at higher frequencies where rheological properties become more important than gravity (Efroimsky 2012a,b).

\section{Limitations on a previously used formula for tidal dissipation} \label{limit.sec}

 Jackson et al. (2008) estimated tidal dissipation in 18 exoplanets, relying on the following expression for the average damping rate:
 \begin{eqnarray}
 \langle\,P\,\rangle~=~\frac{36}{19}~\frac{\pi\,\rho^2\,n^5\,R^7}{\mu\,Q}~e^2~~,
 \label{pc79.eq}
 \label{197}
 \end{eqnarray}
 where $\,\rho\,$ is the mean density, $\,\mu\,$ is the rigidity, and $\,Q\,$ is the tidal quality factor. The formula was
 adopted from the paper by Peale et al. (1979) who referred to their preceding work (Peale \& Cassen 1978). We, however, failed to find an explicit presence of this formula in {\it{Ibid}}.

 In other publications (e.g., Mardling 2007, Murray \& Dermott 1999, Segatz et al. 1988), a different expression is commonly used:
 \begin{eqnarray}
 \langle\,P\,\rangle~=~\frac{21}{2}~\frac{k_2}{Q}~\frac{G~{M^{\,*}}^{\,{2}}~R^5}{a^6}\,n\,e^2~~,
 \label{pc78.eq}
 \label{198}
 \end{eqnarray}
 at times accompanied with a reference to the same paper by Peale \& Cassen (1978). Insertion of the approximate expression
 \be
 k_2~\approx~\frac{3~\rho~\mbox{g}~R}{19~\mu}~~,
 \label{199}
 \ee
 in the equation (\ref{pc79.eq}) transforms the latter into the equation (\ref{pc78.eq}), although with a different numerical factor; namely, with $9$ instead of $21/2$.

 The formula (\ref{pc78.eq}) can be obtained from the equation (31) of Peale \& Cassen (1978). It also ensues from the more general equation (\ref{3}) presented above in our paper, when the following  restrictive assumptions are applied:
 \begin{itemize}
 \item[a.~] the inclination $\,i\,$ of the perturber's orbit on the equator of the perturbed body is set equal to zero;
 \item[b.~] the terms of power 4 and higher in the eccentricity $\,e\,$ are neglected;
 \item[c.~] only quadrupole ($\,l=2\,$) inputs are included;~\footnote{~While $\,l=2\,$ inputs are usually sufficient, sometimes terms with higher values of $\,l\,$ can not be neglected. One such case is Phobos, whose orbital evolution is influenced by the $\,l=3\,$ and, perhaps, even the $\,l=4\,$ terms (Bills et al. 2005). Another class of exceptions is constituted by close binary asteroids. The topic was addressed by Taylor \& Margot (2010), who took into consideration terms up to $\,l=6\,$.}
 \item[d.~] the consideration is limited to bodies rotating $ \, ${\it{synchronously}}$ \, $;
 \end{itemize}
 Under the assumptions [a - c], only the terms with $~(lmpq)\,=\,(201,-1)\,$, $\,(2011)\,$, $\,(220,-1)\,$, $\,(2201)~$ are to be taken into account. From the formula (\ref{moda}), we see that for all these terms the physical forcing frequency $\,\chi_{lmpq}\,\equiv\,|\omega_{lmpq}|\,$ approximately assumes the same value $\,n\,$,
 {{provided the assumption}} [d] is also imposed, i.e., provided that $\,\dot{\theta}\,=\,n\,$. This way, $ \, ${\it{in the case of synchronous spin}}, $\,k_2/Q\,$ assumes the same values for all the four terms taken into account within this approximation.

 Now consider a situation where items [a] and [b] are relaxed, items [c] and [d] are kept, and an extra, highly restrictive item is added:
 \begin{itemize}
 \item[e.~] the Constant Phase Lag (CPL) model of tides is adopted, so the inverse tidal quality factor $\,Q^{-1}_{lmpq}\equiv\,\sin|\,\epsilon_l(\omega_{lmpq})\,|\,$ assumes the same value for all Fourier modes $\,\omega_{lmpq}~$.
 \end{itemize}
 Then the quadrupole part of the dissipated power (\ref{3}) looks as \footnote{~The expression (\ref{e4i4.eq}) is valid for the CPL model (i.e., for a frequency independent $\,k_2/Q\,$). An analogous formula for the CTL model (with $\,k_2/Q\,$ linear in frequency) was written down by Wisdom (2008). Naturally, the higher coefficients in our formulae differ, although the leading terms coincide and contain the same coefficient $\,21/2\,$ as in the expression (\ref{198}) in this section. In Section \ref{eqs.sec}, we shall derive similar formulae for a planet of Maxwell rheology with $\,\tau_{_M}\,\chi\gg 1\,$, in a $\,1:1\,$ spin-orbit state, in a non-resonant rotation, and in a $\,3:2\,$ state.}
 \begin{eqnarray}
 \nonumber
 \langle\,P\,\rangle~=~\frac{k_2}{Q}~\frac{G\,{M^{\,*}}^{\,2}\,R^5}{a^6}~n~\left[\,\left(\frac{3}{2}\,i^2\,-\,\frac{11}{16}\,i^4\right)
 \right.
 \nonumber
 \ea
 \ba
  +\,\left(\frac{21}{2}\,+\,
 \frac{15}{2}\,i^2\,-\,\frac{85}{64}\,i^4\right)\,e^2
 \left.+ \left(\frac{2337}{32}\,+\,\frac{1311}{64}\,i^2\right.\right.
 \nonumber
 \ea
 \ba
  \left.\left.-~\frac{10499}{256}\,i^4\right)\,e^4\,\right]~+~O(i^6)~+~O(e^6)\quad.\quad
 \label{e4i4.eq}
 \end{eqnarray}
 Importantly, for bodies with a significant $\,i\,$ and small $\,e\,$ the term $\,3i^2/2\,$ can be by far greater than the $\,21e^2/2\,$ term (the Earth-Moon system being
 an example). Comparing this with (\ref{198}), we see that the neglect of a finite inclination (or obliquity) is detrimental to the studies of tides in moons and planets with significantly inclined equators, e.g., for the Moon.

 It should be reiterated that the formula (\ref{e4i4.eq}) was obtained under the very restrictive assumptions [d] and [e], i.e., for a planet that is synchronised and whose $\,k_2/Q\equiv k_2\,\sin\epsilon_2\,$ is set frequency independent.


 \section{Case study I: Mercury}\label{merc.sec}

 Of all the planets in the solar system, Mercury is the only one captured into a 3:2 spin-orbit resonance. It is the closest to the Sun and has the largest orbital eccentricity. This makes one wonder if tidal heating could play any role in Mercury's evolution and segregation.

 In the expansion (\ref{3}) for the damping rate, a term numbered with $\,lmpq\,$ contains a multiplier $\,\omega_{lmpq}\,$. For this reason, when the planet is in an $\,lmpq\,$ spin-orbit resonance, the input from the $\,lmpq\,$ Fourier mode into tidal heating is zero. For example, the dominating (at small eccentricities) semidiurnal Fourier tidal mode $\,2200\,$ contributes no heat when the rotator is in the exact 1:1 resonance. The physical meaning of this circumstance is that a Fourier component of the tidal bulge, which moves with the same angular velocity as the perturber, does not lag and, therefore, generates no friction. The other components of the bulge, however, do lag and, thereby, do contribute to heating.

 One exception is the case of a synchronous rotation with $\,e=0\,$, a situation where tidal dissipation ceases completely, the tidal bulge being at rest with respect to the perturbed body. Ultimately, any planet that happens to be a sole companion to its star, should come to this state of complete circularisation and synchronisation, which is the only long-term equilibrium state (Hut 1980, Bambusi \& Haus 2012).

 However, Mercury (as well as several known close-in exoplanets) is a part of a multiple-planet system. The pull from its fellow planets prevents Mercury's
 eccentricity from keeping too low a value. Detailed numerical simulations demonstrate that Mercury's eccentricity has varied over {\ae}ons within a rather wide interval, mostly between 0.1 and 0.3 (Correia \& Laskar 2009), so its current value (0.20563) is not extraordinarily high for this planet. However, this significant eccentricity is not a
 very important factor in the thermal history of Mercury, because in the series (\ref{3}) the leading term (the one with $\,lmpq=2200\,$) is of the order of $\,O(e^0)\,$.

\begin{figure}[htbp]
  \includegraphics[width=8.7cm]{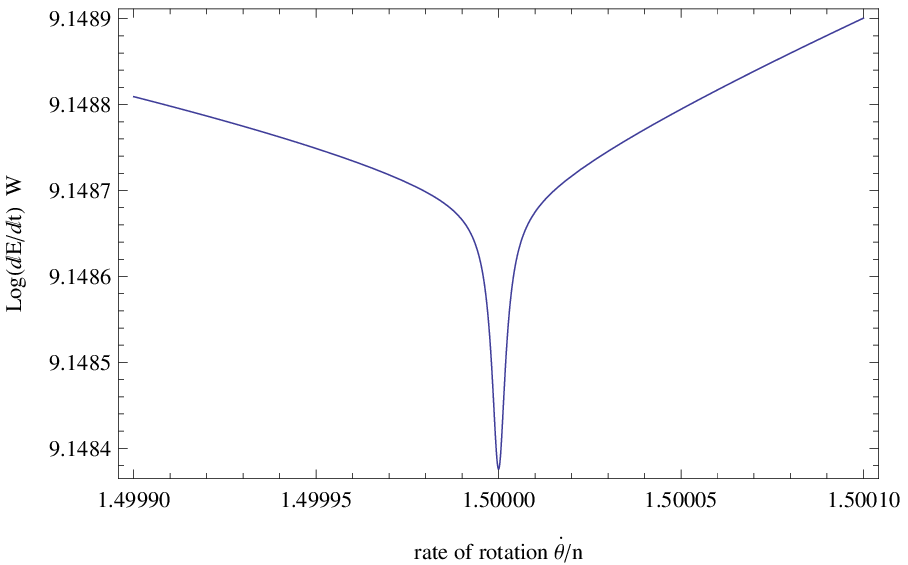}
  \includegraphics[width=8.7cm]{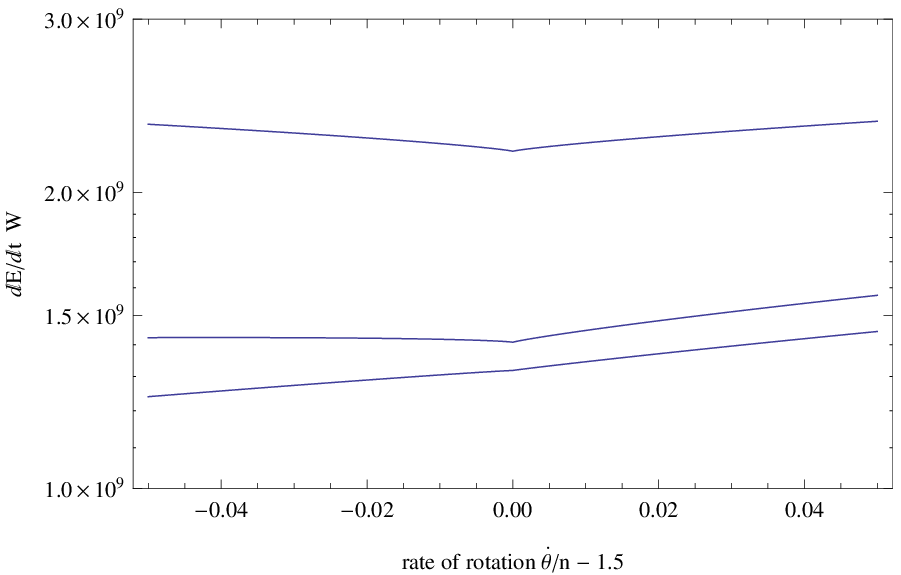}
       \caption{Time-averaged tidal dissipation rate $\,dE/dt\,=\,\langle P\rangle\,$ in a uniform Mercury captured into the 3:2 spin-orbit resonance. Left: decimal logarithm of the
 dissipation rate versus the normalised rotation frequency $\,\dot{\theta}/n\,$, in a close vicinity of the resonance, for $\,e=0.20563\,$. Right: the rate of dissipation versus the normalised rotation rate, for three values
       of eccentricity in the ascending order: $\,e=0.1\,$, $\,e=0.20563\,$, and $\,e=0.3\,$. (The vertical scale in the right pane is log-linear.) \label{mercury.fig}}
       \end{figure}

 Figure \ref{mercury.fig} illustrates the dependence of the damping rate on the dimensionless spin rate $\,\dot\theta/n\,$. The left plot depicts a very narrow vicinity of the resonant frequency, and shows in detail the cleft caused by the vanishing second-largest tidal term $\,lmpq=2201\,$. The cleft is hardly of any practical significance, because the rotation rate of the planet performs forced libration within a much wider range than the one in the graph. The width of this feature is defined mostly by the average viscosity, or by the Maxwell time of the body. The right plot gives the same dependence for a much wider interval of values of the spin rate, and for three values of eccentricity, $\,e=0.1\,$, $\,0.20563\,$, and $\,0.3\,$, going from the lower to the upper curve, respectively. Although increase in the eccentricity yields a stronger dissipation, the dependence is not as strong as in the synchronous-rotation case (cf. Section \ref{io.seq}). In calculating these graphs, we assumed an effective
 rigidity $\,\mu=0.8\cdot 10^{11}\,$ Pa and Maxwell time $\,\tau_{_M}=500\,$ yr, which are close to Earth's values.

 Peale \& Cassen (1978) suggested that the presence of a liquid core inside a planet should enhance tidal damping by roughly 3 -- 15 times, compared to a uniform body of the same mean density and mass. They based this conclusion on the observation that the thinner outer layer (rocky mantle), when supported by a less rigid core, can move more freely under the action of the internal stress. If this conclusion is right, the boost to energy dissipation can be especially strong in Mercury, as its molten core may account for up to 85\% of the total mass. A further increase of the tidal response may come from the possible presence of a solid $\,${\it{Fe$\,$S}}$\,$ layer at the top of the core (Padovan et al. 2014). We would suppose that the actual rate of dissipation can be an order of magnitude higher than what is shown in Figure \ref{mercury.fig}. Even with this upgrade, however, the estimated rate of dissipation is much
 smaller than the production of electric power by the mankind.$ \, $\footnote{~Back in 2012, the world annual electricity net generation was about $\,22500\,$ {\it{TW$ \, $h}}.} It is also very close to the present-day tidal heating rate of the Moon, which is $~\log\,(dE/dt)\,=\,\log\,
 \langle P \rangle
 =9.1~$, the power $\,
  dE/dt\,=\,
 \langle P\rangle
 \,$ being measured in Watts and the logarithm being decimal. So tidal heating is unlikely to have made an impact on the formation of Mercury's molten core.$\,$\footnote{~Qualitatively, our conclusion that tidal heating does not add much to the energy budget agrees with the study by Schubert et al. (1988). In {\it{Ibid.}}, thermal convection lasts for $\,3\,$ Gyr without tidal heating but can, under favourable conditions, be maintained for additional $\,225\,$ Myr if tides are taken into account. Quantitative comparison of our results with those from {\it{Ibid.}} is however impossible, because those authors employed an old model assuming that Mercury formed hot, with early differentiation of the iron core. This is no longer regarded probable -- see, e.g., Noyelles et al. (2014) and references therein.
 \vspace{1mm}\\
 $\left.~~~~\right.$ Our conclusions are in a good agreement with the results obtained by Bills (2002).
 Although Bills claims that tidal damping in Mercury is important, his formulae evidence the opposite. Estimating the tidal damping rate, the author forgot to multiply the overall factor of $\,n^5\,R^5/(2\,G)\,=\,2.49\times 10^{11}\,$ W by the sum of the series itself -- which, very roughly, is of the order of $\,k_2/Q\,$. With that omission corrected, Bills's estimate would become several orders of magnitude lower.}

 \section{Case study II: ~Io}\label{io.seq}

 The most famous manifestation of tidal dissipation is the volcanism of Io. That Io is subject to intense tidal heating was pointed out by Peale et al. (1979) in their cornerstone work which drew considerable attention to the problems of thermal balance in moons. Although the authors brilliantly predicted the semi-molten state of Io's interior, their estimate of damping rate may need re-examination.

 To compute the dissipation intensity, we used our equation (\ref{3}), with Io's inclination set to zero. With the maximal moment of inertia written as $\,C\,=\,\xi MR^2\,$, the coefficient $\,\xi\,$ was assumed to be $\,\xi\,=\,0.37685\,$. As an estimate for the mean rigidity, we adopted a value close to that of the Moon: $\,\mu\,=\,0.65\times 10^{11}\,$ kg m$^{-1}$ s$^{-2}\,$ (Eckhardt 1993). The least-known parameter, the Maxwell time, was set to be $\,\tau_{\rm M}\,=\,1\,$ day, close to the expected value for Titan (F. Nimmo, private communication). The Andrade time, $\tau_{\rm A}$, was set to infinity. Thus, it was assumed that the reaction of the material is purely Maxwell, with no Andrade creep (see the Appendix for details and references). The motivation for the latter decision comes from the fact that Io's mantle is partially molten, so the friction in it is mainly viscoelastic, with no significant input from dislocation unpinning.
  \begin{center}
  \begin{figure}[h]
  \includegraphics[width=3.2in]{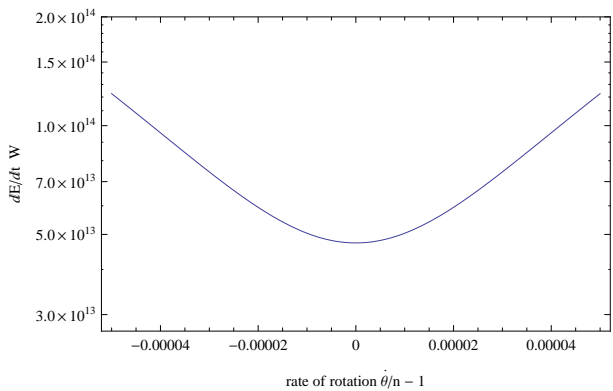}
  \caption{Time-averaged rate of energy dissipation in Io, $\,dE/dt = \langle P\rangle\,$, as a function of the spin rate $\,\stackrel{\bf\centerdot}{\theta\,}\,$, in the vicinity of the 1:1 spin-orbit resonance.}
  \label{io.fig}
  \end{figure}
  \end{center}

 Figure \ref{io.fig} illustrates how the heating depends on the angular velocity $\,\stackrel{\bf\centerdot}{\theta\,}\,$ in the vicinity of the 1:1 spin-orbit resonance. The figure shows the damping rate $\,dE/dt\,=\,\langle P\rangle\,$ plotted against the quantity $~\dot{\theta}/n\,-\,1~$ which is the deviation of the dimensionless spin rate from the synchronous rotation. The synchronous spin is stable, because a slight tilt of the longest axis away from the direction to the planet enables the triaxiality-caused torque to compensate for the time-averaged tidal torque (Goldreich \& Peale 1966, Makarov \& Efroimsky 2013, Williams \& Efroimsky 2012).  As expected on the general dynamical principles, the stable equilibrium (synchronous spin) corresponds to a local minimum of energy dissipation, i.e., to the most energy-frugal position in the considered patch of the phase space. In this minimum, the energy-loss rate is $\,\approx\,5\times10^{13}\,$ W. This is significantly larger than the original estimate by Peale et al. (1979), but is somewhat smaller than the estimate $\,(9.33\,\pm\,1.87)\,\times\,10^{13}\,$ W$ \, $ obtained from astrometric observations by Lainey et al. (2009) who also used an extra assumption that the CPL model is applicable to Io. Given the intrinsic uncertainty of some of our parameters, we find the coincidence up to a factor less than two to be a good match. The fact that the model reproduces (within a factor of two) the result from Lainey et al. (2009) may argue in favour of the Maxwell time being close to one day. For purely Maxwell rheology, the quality factor is inversely proportional to $\tau_{_M}$ if $\tau_{_M}\,n \ll 1$, which is the case here. Therefore, setting $\tau_{_M}=0.1$ days would increase the dissipation rate by a factor of 10. A perfect match with the estimate from observations is achieved for $\tau_{_M}\approx 0.5$ days.

 A word of caution is in order here. Deriving the tidal dissipation rate (\ref{3}), we carried out averaging over one or several periods of tidal flexure. Such a period is not very different from the orbital period. So, by averaging over this timescale, we ignored the contribution from free or forced librations. This approach is legitimate for any long-term state where tidal dissipation is driven mostly by the secular components of polar torque (i.e., anywhere outside spin-orbit resonances). However, in resonances a more accurate treatment would be required, which would bring libration terms into the picture.

  For example, the curve in Figure \ref{io.fig} represents the damping rate that would be achieved if the spin rate stayed at a given near-resonant value. In reality, however, it is only the $ \, ${\it{average}}$ \, $ spin rate that stays resonant, while the $ \, ${\it{instantaneous}}$ \, $ spin rate undergoes  variations over the period of averaging. The planet approaches a spin-orbit resonance relatively slowly, but is captured into resonant rotation very quickly, typically within one period of free libration (e.g., Makarov 2013). In the process of capture into a resonance $\,(2+q):2\;$, the evolution of the angle $~\gamma\,\equiv\,\theta\,-\,\left(1\,+\,{q}/{2}\,\right)\,{\cal{M}}~$ abruptly switches from circulation to oscillation, and the orbit-average spin rate $\,\stackrel{\bf\centerdot}{\theta\,}\,$ assumes a near-resonant value. Immediately after the capture, the magnitude of free librations is close to the maximal possible value, but these librations are quickly damped by tidal friction. However, the forced librations do not go away because they are caused by the eccentricity. As a result, the instantaneous spin rate oscillates around the resonant value, insofar as the neighbouring planets's gravity keeps the residual eccentricity nonvanishing.

  To understand the importance of the libration terms of $\,
  \stackrel{\bf{\centerdot}}{\theta\,}$, note that, generally, these terms multiplied by the harmonics of torque do not average out to zero. So libration contributes to the power exerted by the tidal torque and, thereby, to the dissipation. Therefore, in the presence of librations, the energy dissipation rate is $ \, ${\it{higher}}$ \, $ than it would have been without libration. Unfortunately,
  the Fourier decomposition of the tidal and triaxial torque is very complex, both for the Andrade model and for its simplified version, the Maxwell model. It is not obvious whether a satisfactory analytical treatment of the problem can be obtained. For now, we resort to an approximate, qualitative reasoning described below.

 To estimate the role of physical librations in heating, we simulated the spin of Io subject to both the triaxiality-caused torque and the tidal torque, whose averages balance one another and make the synchronous rotation state that of a stable equilibrium. The formulae for these torques can be looked up in our preceding paper (Makarov et al. 2012, equations 4 - 6). The simulation demonstrates that the forced libration of Io ranges, approximately, from
 $\,-0.5\,$ to $\,0.3\,$ arcsec in the libration angle $\,\theta \,-\,{\cal M}\,$, and within $\,\pm\,2\,\times\,10^{-6}\,$ in $\,\dot\theta/n\,-\,1\,$. Assuming that there are no free or other long-period librations present, the tiny amplitude of the forced librations samples a tiny segment around the minimum of the $\,dE/dt\,=\,\langle P\rangle\,$ curve in Figure \ref{io.fig}. Within that vicinity, the curve is quite flat, and the variation of dissipation rate due to libration is negligible. However, the amplitude of the forced librations is sensitive to the triaxiality parameter $\,(B-A)/C\,$ (and, of course, to the eccentricity $\,e\,$). In our calculation, we used the value $\,(B-A)/C=6.4\times 10^{-3}\,$ borrowed from Anderson et al. (2001). If we increase $\,(B-A)/C\,$ by a factor of two, we shall find the half-amplitude of libration to increase to $\,\approx\,4\,\times \,10^{-5}$. Due to the concavity of the $\,dE/dt\,=\,\langle P\rangle\,$ curve, the rate of dissipation goes up by roughly a factor of two. We see that the shape of a moon plays a significant role in its tidal heating.

 We conclude that, with the other parameters equal, less axially-symmetric (more triaxial) moons should be subject to a significantly stronger heating than their more rotationally symmetric peers. Io represents a borderline case, obviously being close to complete meltdown. It appears entirely plausible that Io had a more elongated shape in the past. Later, because of the excessive tidal heating, it melted down (or, rather, up) to the surface and underwent a drastic reshaping. Acquiring a more symmetric shape helps a tidally perturbed body to lower the heat production in the state of synchronous rotation. The diminished heat flux allows the crust to emerge. The upper mantle becomes colder and less prone to alter its shape under varying tidal stresses. So the tidal bulge solidifies and becomes the new triaxial figure. Speculatively, Io could have gone through several such seesaw variations, having gradually reshaped itself to more symmetrical forms, especially if the rise of dissipation was assisted by episodical boosts in eccentricity or inclination.

 The above reasoning is qualitative, so it requires further numerical confirmation. Results of numerical modeling of this situation will be reported elsewhere.$\,$\footnote{~The influence of librations upon tidal heating of Enceladus was studied analytically by Wisdom (2004). He considered a very special case where the
 libration period was about three times longer than the orbital period, so the direct employment of the formula for the time-averaged damping rate was legitimate,
 at least for qualitative estimates. Also note that in {\it{Ibid}} the CTL (constant time lag) model was used.}

 \section{Case study III: ~Kepler-10$ \, $b}\label{case3}

 Kepler-10b was the first confirmed terrestrial planet discovered outside the Solar System (Batalha et al. 2011). It is located remarkably close to its host star, the semimajor axis being only $\,2.520\times 10^9\,$ m, which is less than 0.017 AU. Among the super-Earths discovered with the sensitive Kepler photometer, Kepler-10$ \, $b$ \, $ stands out as one of the smallest and densest bodies known outside the Solar system. With an estimated mass of $\,4.44\,M_{earth}\,$ and the radius $\,1.42\,R_{earth}~$ ({\it{Ibid.}}), the mean density of the planet comes up to $\,8640\,$ kg m$^{-3}$, which is almost 60\% greater than the mean density of the Earth, the densest planet in the solar system. While the remarkable fact that the Earth is four to five times denser than Jupiter was known already to Sir Isaac Newton (1687), here we are dealing with a planet considerably more massive than the Earth and several
 times more dense than gas giants. This leaves little doubt that the planet is terrestrial, unlike the distinct category of ``hot Jupiters" which are more massive but have mean densities between 0.3 and 3 densities of Jupiter. The mean density of the Earth interior is equal to the local density at approximately $\,3500\,$ km radius, where the core-mantle boundary is located. The greater density of Kepler-10$ \, $b$ \, $ may very well indicate that the relative radius of its molten core (the actual radius of the core, divided by the overall radius of the planet) is larger than the relative radius of the molten core of the Earth. If this is the case, then Kepler-10$ \, $b$ \, $ may be classified, in terms of its internal composition, as a super-massive Mercury.$\,$\footnote{~It should be noted that our understanding of terrestrial exoplanets does not stand only on comparisons with the density of the Earth, as the compressibility of the mantle has to be taken into account for the large pressures reached inside massive planets. Various works have addressed the possible internal structure of these objects in general and of Kepler-10b in particular (e.g., Grasset et al. 2009, Valencia et al. 2010, Zeng \& Sasselov 2013).}
 Following Peale \& Cassen (1978), we speculate that the core can boost tidal damping by a factor of a few to several.
 However, we shall not attempt to take this extra boost into account, because it is not large enough to change our conclusions.

 \subsection{The spin state, orbit motion and rheology. Educated guess\label{guess}}


 Presently, we possess observational data neither on the rotation of Kepler-10$ \, $b$ \, $, nor on its obliquity. The eccentricity of Kepler-10$ \, $b$ \, $ could not be determined
 in Batalha et al. (2011), because the signal detected in the follow-up spectroscopic observations of the host star was too weak for a confident estimation. A recent analysis carried out by Fogtmann et al. (2014) indicates that the eccentricity is extremely small. Although the value $~0.050^{+0.012}_{-0.050}~$ provided in {\it{Ibid}}. is consistent with the eccentricity being zero, it should be interpreted as an upper limit. Setting $\,e=0\,$ is not an option, because the orbit is likely to be
excited by a more massive neighbour, the planet Kepler-10c.

 Under regular circumstances, tidal dissipation of the orbital kinetic energy in a two-body system is wont to damp both the eccentricity and obliquity. Important exceptions are:
 \begin{itemize}
 \item[~\it 1.~~] Multiple-planet systems, where mutual interactions between the planets can pump up both the eccentricity and obliquity of the inner planet (Correia et al. 2012; Greenberg et al. 2013).
 \item[~\it 2.~~] Situations where either a close-in planet or the star rotates faster than the orbital motion in the prograde sense. In particular, if the star rotates faster than $\,n\,$, the tidal bulge on it leads the direction to the planet. An increase in both $\,e\,$ and $\,a\,$ ensues (see, e.g., Murray and Dermott 1999). The lag on the star may be small, but it is enough to keep the planet's eccentricity nonvanishing. A slow tidal dissipation in the star also means it can retain its fast rotation for a long time, no matter how massive the close-in planets happen to be. The described situation is analogous to the Earth-Moon system whose eccentricity and semi-major axis are both increasing.
 \end{itemize}
 Thus, finite residual eccentricities and obliquities should not be unusual for close-in planets. The presence of the more massive and distant planet Kepler-10$ \, $c$ \, $ with an orbital period of 43.3 days \citep{fre,dum}, makes it likely that the inner planet is neither completely circularised nor aligned. So we consider small residual values of $\,e\,$ and $\,i\,$. Somewhat arbitrarily, we chose two cases: one of $\,e=0.001\,$ and $\,i=0.001\,$, another of $\,e=0.001\,$ and $\,i=0.0001\,$. However, the possibility of larger values cannot be precluded.

 For the close-in super-Earths GJ 581$ \, $d$ \, $ and GJ 667$ \, $Cc$ \, $, which are members of multiple systems, a 3:2 or higher spin-orbit resonance was found to be a more likely end-state than the synchronous rotation, provided the initial spin rate was high in the prograde sense (Makarov et al. 2012; Makarov \& Berghea 2013). For Kepler-10$ \, $b$ \, $, however, tidal interactions are stronger; so the chances of this overheated (and, possibly, semi-molten) planet being in a higher than synchronous spin-orbit resonance are far from obvious, as we shall see shortly.

 The next most significant uncertainty in our analysis is the rheology of Kepler-10$ \, $b$ \, $. The frequency dependence
of $\,{k}_{\textstyle{_2}}/Q\,$ is defined by two major physical circumstances, the self-gravitation of the planet and the rheology of its mantle. A rheological law (i.e., an equation interconnecting the strain and the stress) contains contributions from elasticity, viscosity and inelastic processes (mainly, dislocation unjamming). Together, these three factors render a so-called Andrade creep (Efroimsky 2012$ \, $a, 2012$ \, $b). It should be noted that a mantle behaves as the Andrade body at higher frequencies only, and changes its behaviour toward the Maxwell model at lower frequencies. This happens because, at frequencies below a certain threshold, only elasticity and viscosity contribute to the rheological response of the mantle. Above the threshold, dislocation unpinning (unjamming) plays a considerable role. The value of the threshold frequency is highly sensitive to the temperature of the mantle, as can be seen from formula (17) in Karato and Spetzler (1990). The formula indicates that, for realistic binding energies, a 10 to 20~\% increase in temperature can increase the threshold frequency by an order or two of magnitude. Given that for the Earth the threshold is of the order of 1 yr$^{-1}\,$, we see that for overheated planets the threshold may be as high as 1 day$^{-1}\,$. It would be even higher for higher temperatures of the mantle.

 Speaking of the planet Kepler-10$ \, $b$ \, $, we assume that, owing to intensive tidal heating, its mantle should contain a lot of partial melt and thus have a low average viscosity. The Maxwell time, therefore, is likely to be much shorter than those of the Earth or Mercury. It should be closer to the Maxwell times for icy satellites, which is believed to be of the order of days.
 %
 %
 With an orbital period about one day, Kepler-10$ \, $b$ \, $ should experience tides at frequencies of the order 1 day$^{-1}\,$, these frequencies likely being below the Andrade-Maxwell threshold. So the Andrade mechanism of tidal friction (unpinning of dislocations) is likely to be less significant for this planet, allowing us to use a purely Maxwell model.$\,$\footnote{~In the past, several other rheological models were employed in the literature (e.g., Henning et al. 2009, Heller et al. 2011, Henning \& Hurford 2014).} Armed with these considerations, we now have to build the so-called quality functions $~k_l(\omega_{\textstyle{_{lmpq}}})~\sin\epsilon_l(\omega_{\textstyle{_{lmpq}}})~$ standing in the expression (\ref{3}) for the damping rate.

 \begin{center}
 \begin{figure}[h]
 \includegraphics[width=3.4in]{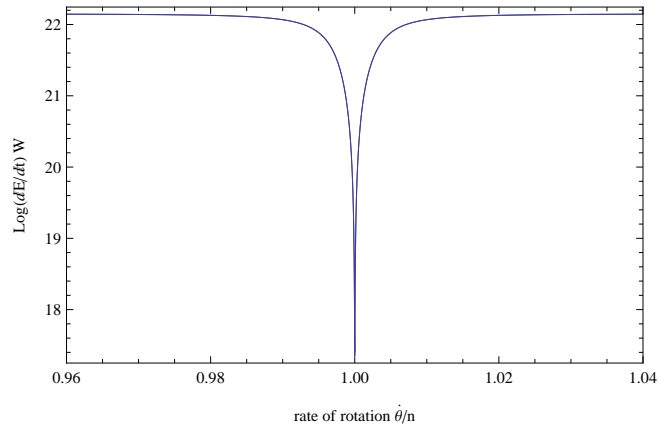}\\
 \caption{Time-averaged rate of energy dissipation $\,dE/dt\,=\,\langle P\rangle\,$ in Kepler-10$ \, $b, as a function of the dimensionless rotation rate $\,\dot{\theta}/n\,$,
 in the vicinity of the 1:1 spin-orbit resonance. The two curves (one computed for $\,\tau_{_M}=10\,$ days, $\,e=0.001\,$,  $\,i=0.001\,$, another for $\,\tau_{_M}=10\,$ days, $\,e=0.001\,$, $\,i=0.0001\,$) virtually coincide and can barely be distinguished from one another.}
 \label{kepler.fig}
 \end{figure}
 \end{center}

 \subsection{Tidal dissipation rate in the 1:1 spin-orbit resonance}

Each term of the series (\ref{3}) contains a quality function. These are calculated by the below formula (\ref{fora}), with the expression (\ref{dora}) built in. The result is presented in Figure \ref{kepler.fig} which depicts the dependence of tidal damping upon the spin rate of Kepler-10$\,$b$\,$ (assuming it has a rocky mantle).
For this computation, however, we assumed a rather short Maxwell time of 10 days, taking into account that the mantle may have a
lot of partial melt in it. A small residual eccentricity of 0.001 was also accepted, and two values of $\,i\,$ were explored: 0.001 and 0.0001.

 The curves corresponding to the two values of inclination are so close on the graph that it is difficult to see a separation between them. We also note that everywhere outside a narrow vicinity of the 1:1 resonance the rate of damping is flat, i.e., almost independent of the spin rate. The sharp cleft is easily explained by the expression (\ref{fora}) from which we see that $\,k_2/Q\,$ vanishes in the zero-frequency limit. More generally, an $\,lmpq\,$ term of the series (\ref{3}) vanishes when the tidal mode $\,\omega_{\textstyle{_{lmpq}}}\,$
 goes through zero, while outside the resonance the input from this term is relatively flat. More subtle variations of
 tidal dissipation rate around the resonance are concealed in this figure by the logarithmic scale.

 Both the perceived flatness of the curve outside the main resonances and the apparent weak dependence on $\,i\,$ are explained by the approximate equation (\ref{k11.eq}) derived in Section \ref{eqs.sec}, for the special case of $\,\tau_{_M}\,n \gg 1\,$, which is valid for the chosen parameters of Kepler-10$\,$b, as well as for a large class of short-period super-Earths that are not completely molten. If we fix the obliquity at $\,i=10^{-5}\,$, we obtain the following estimates for the rate of dissipation at the exact 1:1 resonance: $\,\log(dE/dt)=17.30\,$, $\,15.36\,$, and $\,13.30\,$ for $\,e=10^{-3}\,$, $\,10^{-4}\,$, and $\,10^{-5}\,$, respectively. The rate of dissipation increases by almost exactly two orders of magnitude for each order of magnitude increase in $\,e\,$, as expected from the equation (\ref{k11.eq}) when the $\,O(i^2)\,$ terms in it are small. In the tidal regime in question, when $\,\tau_{_M}\,n \gg 1\,$, the quality function is proportional to $\,\tau_{_M}\,$, as can be seen from the equation (\ref{kosa}). Therefore, the dissipation rate is less strongly dependent on $\tau_{_M}$ than on $e$, and the inevitable uncertainty in the former
parameter is relatively less restrictive. The absolute values in Figure \ref{kepler.fig} can be used only for very general
guidance and comparison with the previous estimates for Mercury and Io, but the character of the curves is valid
for a significant range of these critical parameters.

 Thus, in synchronous spin-orbit resonance, Kepler-10$\,$b$\,$ will dissipate less energy, by roughly five orders of magnitude, than in any other rotation state, including the 1:2 and 3:2 resonances. The ensuing implications for the destiny of such close-in planets are dramatic. If a planet does not succeed in falling into the state 1:1, and gets captured into a higher spin-orbit resonance, the rate of tidal dissipation in the planet becomes so high that its temperature should be growing by several degrees per year.$\,$\footnote{~For the planet's heat capacity we adopted a value of $\,1200~\mbox{J}~\mbox{kg}^{-1}~\mbox{K}^{-1}\,$ from B\v{e}hounkov\'a et al. (2011).}  This
 should be enough to quickly melt the planet to the surface and make it a ball of magma. In a very close vicinity of the host star, planets rotating synchronously may remain solid for a longer time than asynchronous planets. Still, even synchronised planets may not be able to survive for longer than $\sim 1$ Myr in a solid form. The existence of close-in, high-density planets requires scenarios of their survival at a higher level of complexity, which remain somewhat speculative because of the lack of accurate data.

 \subsection{Possible scenarios for extremely close-in terrestrial planets}

 One possible scenario for a close-in terrestrial planet is the following. If the orbital eccentricity and obliquity are not excited by a third body, and the star does not pump up these parameters by the transfer of angular momentum from its own rotation, the orbit should relatively quickly circularise, and the obliquity should decrease. This would drive the tidal dissipation down to small values. As we explained above, in the space of parameters there exists a dip wherein the tidal dissipation rate is minimal. This is the synchronous rotation with a zero or near-zero obliquity. In this regime, the damping rate is by orders of magnitude lower than in a non-resonant state or in a higher resonance. In the presence of a non-zero residual eccentricity, the planet should also be almost perfectly spherical in order to get a respite from the excessive tidal heating through libration (see Section \ref{io.seq}).

 In multiple systems, the eccentricity and obliquity of close-in planets can be excited by external interactions. In this situation, a young planet gets completely molten even if it is synchronised -- so it loses its permanent figure before the orbit circularisation and obliquity decrease take place. Residing at the bottom of the energy dissipation dip ($\,e\approx 0\,$, $\,i\approx 0\,$, spin = 1:1), the planet then begins to cool down and may eventually solidify on the surface. The stationary tidal bulge becomes the permanent figure of the newly formed mantle. But the planet is safe now, sitting in the dip and dissipating almost no energy due to its more axially symmetric shape. The tidal evolution of the orbit and obliquity ceases too, unless the tidal dissipation in the star can drive the eccentricity to higher values again.

 \section{Analytic approximations for a warm Maxwell planet}\label{eqs.sec}

 Introduced as a function of the Fourier tidal mode $\,\omega_{\textstyle{_{lmpq}}}\,$, the product $\,k_l(\omega_{\textstyle{_{lmpq}}})~\sin\epsilon_l(\omega_{\textstyle{_{lmpq}}})~$ can be also written down as a function of the positive definite forcing frequency $\,\chi_{\textstyle{_{lmpq}}}\,\equiv\,|\,\omega_{\textstyle{_{lmpq}}}\,|\,$:
 \begin{eqnarray}
 \nonumber
 k_l(\omega_{\textstyle{_{lmpq}}})~\sin\epsilon_l(\omega_{\textstyle{_{lmpq}}})\qquad\qquad\qquad\qquad\qquad\nonumber\\
 \nonumber\\
 \nonumber
 =~k_l(\chi_{\textstyle{_{lmpq}}})~\sin\epsilon_l(\chi_{\textstyle{_{lmpq}}})~\,\mbox{Sgn}\,\omega_{\textstyle{_{lmpq}}}\\
 \nonumber\\
 ~=~\frac{k_l(\chi_{\textstyle{_{lmpq}}})}{Q_l(\chi_{\textstyle{_{lmpq}}})}~\,\mbox{Sgn}\,\omega_{\textstyle{_{lmpq}}}~~,~~~
 \label{}
 \end{eqnarray}
 see Section \ref{brief}.

 For a homogeneous planet obeying the Maxwell rheological law, the frequency dependence of $\,{k}_{\textstyle{_l}}/Q_l={k}_{\textstyle{_l}}(\chi)\,\sin\epsilon_{\textstyle{_l}}(\chi)\,$ is furnished by the expression
 \begin{eqnarray}
 ^{\textstyle{^{(Maxwell)}}}{k}_{\textstyle{_l}}(\chi)\;\sin\epsilon_{\textstyle{_l}}(\chi)\qquad\qquad\qquad\qquad\qquad\nonumber\\
 \nonumber\\
 ~=~\frac{3}{2}\;\,\frac{{\cal{A}}_{\textstyle{_l}}\;(\chi\,\tau_{_M})^{-1}}{
 \,\left(\;1\;+\;{\cal{A}}_{\textstyle{_l}}\;\right)^2\;+\;(\chi\,\tau_{_M})^{-2}\,} ~~~,~~~~~
 \label{fora}
 \end{eqnarray}
 derived in the Appendix. $ \, $Here $\,\tau_{_M}\,$ is the Maxwell time, $\,\chi\,=\,\chi_{lmpq}\equiv|\omega_{lmpq}|\,$ is the
 forcing frequency corresponding to an $\,lmpq\,$ tidal mode, and  $\,{\cal{A}}_{\textstyle{_l}}\,$ are dimensionless factors reflecting the interplay of
 self-gravitation and rheology in tidal response. Being interested in the principal (quadrupole) part of the expansion (\ref{3}), we need the expression
 for $\,{\cal{A}}_{\textstyle{_2}}\,$:
 \begin{eqnarray}
 {\cal{A}}_{\textstyle{_2}}&\equiv&\frac{\textstyle{19~\mu}}{\textstyle{2\,\mbox{g}\,\rho\,R}}~=~\frac{19}{2}~\frac{\mu~R}{G\,\rho\,M}\qquad\qquad\qquad
  \nonumber\\
 \nonumber\\
 &=&\frac{\textstyle{57~\mu}}{\textstyle{8\,\pi\,G\,\rho^2\,R^2}}~
   ~=~\frac{\textstyle{57}}{\textstyle{8}\,\pi\,G\,\rho^2\,R^2~\,J}~~~,\qquad~~
 \label{dora}
 \end{eqnarray}
 $\rho$, g, $R$, $M$ being the planet's mean density, surface gravity, radius and mass; $G$ being the Newton gravitational constant; and $\mu$ and $J=1/\mu$ being the unrelaxed rigidity and compliance, respectively. For an Earth-sized planet, $\,{\cal{A}}_{\textstyle{_2}}\approx 2\,$ (Efroimsky 2012$\,$b, $\,$Table 1).

 By inserting the expression (\ref{dora}) into (\ref{fora}), plugging the result into the series (\ref{3}), retaining only the $\,l=2\,$ part, and expanding it over $\,e\,$ and $\,i\,$, we arrive at an expression for the dissipation rate, written as a series over powers of $\,e\,$ and $\,i\,$. In the special case of a warm but not completely molten super-Earth or icy satellite, we can simplify the series further by assuming that $\,\tau_{_M}\,\chi\,\gg\,1\,$. With this simplification taken into account, and after truncating powers six and higher, we obtain an approximation for the time-averaged energy-damping rate $\,dE/dt\,=\,\langle P\rangle\,$ in a synchronised planet:
 \begin{eqnarray}
 \nonumber
 \langle\,P\,\rangle
 \,=~\frac{3}{2}~\frac{{G}{M^{\,*}}^2 \,R^5}{a^6}\frac{{\cal{A}}_2}{\tau_{_M} (1+{\cal{A}}_2)^2}
 \left[\left(\frac{3}{2}\,i^2\right.\left.-\,\frac{37}{32}\,i^4\right)\right.
 \nonumber
 \ea
 \ba
 \nonumber
 \left.\,+\,\left(\frac{21}{2}\,-\,
 \frac{27}{8}\,i^2\,+\,\frac{27}{16}\,i^4\right)\,e^2 \right.
 \left.+ \left(\frac{1125}{64}\,+\,\frac{213}{32}\,i^2\right.\right.
 \nonumber
 \ea
 \ba
 \left.\left. -~\frac{3499}{256}\,i^4\right)\,e^4\,\right]~+~O(e^6)~+~O(i^6)\quad,\qquad\qquad
 \label{k11.eq}
 \end{eqnarray}
 and in a non-resonant planet:
 \begin{eqnarray}
 \nonumber
 \langle P\rangle =\frac{3}{2}\,\frac{{G}{M^{\,*}}^2 \,R^5}{a^6}\,\frac{{\cal{A}}_2}{\tau_{_M} (1+{\cal{A}}_2)^2}
 \left[\left(\frac{3}{4}+\frac{3}{4}\,i^2-\,\frac{13}{16}i^4\right)
 \right.
 \nonumber
 \ea
 \ba
 \nonumber
 \left.
  \,+\,\left(\frac{27}{4}\,+\,
 \frac{9}{4}\,i^2\,-\,\frac{39}{16}\,i^4\right)\,e^2\right.
 \left.+ \left(\frac{1503}{64}\,+\,\frac{9}{2}\,i^2
 \right.\right.
 \nonumber
 \ea
 \ba
 \left.
 \left.
 -\,\frac{3849}{256}\,i^4\right)\,e^4\,\right]~+~O(e^6)~+~O(i^6)\quad,\qquad\qquad
 \label{knone.eq}
 \end{eqnarray}
 and in a planet trapped in the 3:2 resonance:
 \begin{eqnarray}
 \nonumber
 \langle P \rangle = \frac{3}{2}\,\frac{{G}{M^{\,*}}^2 \,R^5}{a^6}\,\frac{{\cal{A}}_2}{\tau_{_M} (1+{\cal{A}}_2)^2}
 \left[\left(\frac{3}{4}+\frac{3}{4}i^2-\frac{13}{16}i^4\right)\right.
 \ea
 \ba
 \nonumber
  +\left.\left(-\frac{39}{16}\,+\,
 \frac{183}{16}\,i^2-\,\frac{851}{128}\,i^4\right)e^2+ \left(\frac{2043}{32}\,-\,\frac{2295}{64}\,i^2\right.\right.
 \nonumber
 \ea
 \ba
 \left.\left.+\;\frac{1773}{512}\;i^4\right)\,e^4\,\right]~+~O(e^6)~+~O(i^6)\quad,\qquad
 \label{k32.eq}
 \end{eqnarray}
 where $\,M^{\,*}\,$ is the mass of the star. As ever, $\,P\,$ is the power exerted by the tidal stresses, and $\,\langle\,.\,.\,.\,\rangle\,$ denotes time averaging over one or several cycles of tidal flexure.
 Insofar as the truncation of $\,O(e^6)+O(i^6)\,$ is legitimate
 (conservatively, for $e\la 0.2$), three conclusions stem from the above formulae.
  \begin{itemize}
  \item[~1.~] In synchronised planets, the leading-order inputs into the energy dissipation rate $\,dE/dt\,=\,\langle P\rangle\,$ must scale as $\,3/2\,i^2\,$ and $\,21/2\,e^2\,$. Accordingly, $\,\langle\,P\,\rangle\,$ in such planets scales as either $\,3/2\,i^2\,$ or $\,21/2\,e^2\,$, whichever is greater.
  \item[~2.~] Tidal dissipation in non-resonant planets is virtually independent of $\,e\,$ or $\,i\,$.
  \item[~3.~] Likewise, the dissipation rate at the 3:2 resonance is virtually independent of $\,e\,$ or $\,i\,$.
  \end{itemize}
 The latter conclusion may look somewhat counterintuitive, but it is easily propped up by the following observation. In the series (\ref{3}) for the damping rate, the semidiurnal ($lmpq=2200$) term is the largest and it scales with both $\,e\,$ and $\,i\,$ as $\,O(1)\,$. The second-largest term (the one with $lmpq=2201$) turns out to be proportional to $\,3n-2\dot{\theta}\,$, whereby it vanishes in the 3:2 spin-orbit resonance. Hence, in this resonance, we are left with the obliquity- and eccentricity-independent
 semidiurnal term, as well as many terms that are much smaller. In Figure \ref{kepler.fig}, the two curves (corresponding to the case of $\,e=0.001\,$, $\,i=0.001\,$ and to that of $\,e=0.001\,$, $\,i=0.0001\,$) virtually coincide, because in the equation (\ref{k11.eq}) the dominating term scales as $\,21/2\,e^2\,$, the obliquity-dependent terms being less important.

 \section{Conclusions}

  We have demonstrated that tidal dissipation is considerably more involved a topic than was assumed in many studies conducted after the seminal work by Peale \& Cassen (1978).  The commonly accepted in the literature approximate formula (\ref{pc78.eq}) for the damping rate follows from the equation (31) in Peale and Cassen (1978), provided that the inclination (or obliquity) is set zero and higher-order terms in the eccentricity are neglected. It can also be derived from a more general expression, our formula (\ref{3}), under an extra assumption that the rotator is synchronised.

On the examples of Mercury, Io, and Kepler-10$\,$b, we addressed a broad range of issues emerging from the so-revised theory
of tidal dissipation. The main practical highlights are:

 1.  Like Mercury, close-in exoplanets of terrestrial composition may be captured into stable, long-term asynchronous resonances, such as 3:2 or 2:1. In such states, the planets have a net rotation with respect to the mean direction to the star. The tidal bulge runs across their surface, which results in a dissipation rate that is higher, by orders of magnitude, than the dissipation rate in a synchronised planet. This conclusion is fortified by our expressions (\ref{k11.eq}), (\ref{knone.eq}), and (\ref{k32.eq}) for the damping rate in a planet, in the cases when it is synchronised, or nonresonant, or in a $\,3:2\,$ spin-orbit resonance, respectively. These formulae were derived for a planet that is described with the Maxwell rheology and is sufficiently close-in (so that $\,\tau_{_M}\,\chi\,\gg\,1\,$, where $\,\tau_{_M}\,$ is the Maxwell time and $\,\chi\,$ is the principal tidal frequency).

 2. Planet-planet orbital interactions play a crucial role in defining the ultimate fate of those rocky planets that managed to get close to their stars. If a considerable eccentricity is secularly excited by the outer companions, both the orbital evolution rate and the tidal heating become boosted by a few to several orders of magnitude. Our preliminary calculations show that such planets should be liquefied, even when they are settled in the absolute energy minimum (the 1:1 resonance, with a zero or near-zero inclination).

 A planet can, however, survive in the rocky state, provided there is no significant planet-planet orbital interaction pumping up its eccentricity or the obliquity. For such survivors, the tidal dissipation in the host star may become an important factor. Specifically, if the rotation of the star is prograde and is faster than the orbital motion, it will pump up the eccentricity and may also lead to a finite obliquity that, in turn, will perturb the orbit inclination (Teyssandier et al. 2013). All these circumstances will channel the kinetic energy into the heating of the close-in planet, resulting in its liquefaction. It appears that most of the host stars with transiting close-in giant exoplanets rotate slower than these planets' $\,n\,$ (Matsumura et al. 2010, Table 1).

 3.  We have hypothesised that the tidal damping rate can be considerably boosted by physical librations. The hypothesis stems from the following considerations.
 An $\,lmpq\,$ term of the expression (\ref{3}) for the damping rate contains a multiplier $\,\,k_l(\chi_{\textstyle{_{lmpq}}})\,\sin\epsilon_l(\chi_{\textstyle{_{lmpq}}})\,$ that depends on the physical frequency $\,\chi_{\textstyle{_{lmpq}}}\,$.
    This dependence is extremely sharp near resonances, i.e., in closest vicinities of the zeroes of the frequency. As obvious from the expression (\ref{2}) for the frequency, we can interpret the
    multipliers $\,k_l(\chi_{\textstyle{_{lmpq}}})\,\sin\epsilon_l(\chi_{\textstyle{_{lmpq}}})\,$ as functions of the rotation rate $\,\stackrel{\bf\centerdot}{\theta\,}$. Their dependence on $\,\stackrel{\bf\centerdot}{\theta\,}$ will also be very sharp when a resonance is near (i.e., when $\,\stackrel{\bf\centerdot}{\theta\,}$ is very close to $\,(l-2p+q)~n/m\,~$).
    Due to the sharp form of this dependence, even a tiny deviation of $\,\stackrel{\bf\centerdot}{\theta\,}$  from a resonant value will change the effective value of $\,k_2/Q\,$
considerably. This situation is best illustrated by Figure \ref{kepler.fig}, where the dependence of
the average dissipation rate upon $\,\stackrel{\bf\centerdot}{\theta\,}$  is depicted in a close vicinity of the 1:1 spin-orbit resonance.

   The sensitivity of the energy damping rate to the values of $\,\stackrel{\bf\centerdot}{\theta\,}$  indicates the key role played by the physical libration in the tidal heating process. Although physical libration does not change the $\,${\it{mean}}$\,$ value of the spin rate (which
   stays resonant), the libration yields variations of the $\,${\it{instantaneous}}$\,$ value of $\,\dot{\theta}\;$. We have provided qualitative argumentation showing that these variations should increase the overall rate of heat production. However, our physical arguments are not yet rigorous proof. The latter needs to be obtained through accurate numerical simulations.

 4. The magnitude of libration in the spin rate being defined by the planet's triaxiality, the latter should be a significant factor determining the dissipation rate at spin-orbit resonances. Other parameters being equal, a body with a more pronounced triaxiality should generate more heat than a similar body of a more symmetrical shape. On the other hand, we surmise that a feedback may also exist, in that the rate of tidal heating may change the shape of close-in planets through repeated episodes of complete melt-down.

 \section*{Acknowledgments}

\noindent
 The authors deeply thank both referees (Patrick Taylor and an anonymous referee) for their detailed and very thoughtful report on earlier versions of this work. The authors are also indebted to James G. Williams for reading the manuscript and offering very important comments. All these colleagues have helped the authors greatly to improve the quality of the paper.



 \appendix
 \begin{center}
  {\underline{\Large{\bf{Appendix}}}}
 \end{center}

 \section{How rheology and self-gravitation determine the\\ frequency dependencies of Love numbers and phase lags}\label{rheology}\label{first}

 The time-averaged dissipation rate in a homogeneous planet is given by the expression (\ref{3}), provided the apsidal precession of the star, as seen from the planet, is uniform. An $\,lmpq\,$ term of that expression contains a {{quality function}} $\,\,k_l(\omega_{\textstyle{_{lmpq}}})\,\sin\epsilon_l(\omega_{\textstyle{_{lmpq}}})\,$. $\,$Interplay of self-gravitation and rheological properties of the planet makes the forms of these functions
 nontrivial, although some qualitative features of these dependencies are generic and invariant of rheology and size.

  As demonstrated, e.g., in Efroimsky \& Makarov (2014), a quality function of a Fourier mode $\,\omega_{\textstyle{_{lmpq}}}\,$ can always be written down as a function of the appropriate physical frequency $\,\chi_{\textstyle{_{lmpq}}}\,=\,|\,\omega_{\textstyle{_{lmpq}}}\,|\,~$:
 \begin{eqnarray}
 k_l(\omega_{\textstyle{_{lmpq}}})~\sin\epsilon_l(\omega_{\textstyle{_{lmpq}}})~=~k_l(\chi_{\textstyle{_{lmpq}}})~\sin\epsilon_l(\chi_{\textstyle{_{lmpq}}})~\,\mbox{Sgn}\,\omega_{\textstyle{_{lmpq}}}~~~.
 \label{}
 \end{eqnarray}
 The following was derived in Efroimsky (2012$ \, $a,$ \, $b) for a homogeneous spherical body:
 \begin{eqnarray}
 {k}_{\textstyle{_l}}(\chi)\;\sin\epsilon_{\textstyle{_l}}(\chi)\;=\;\frac{3}{2\,({l}\,-\,1)}\;\,\frac{-\;{\cal{A}}_{\textstyle{_l}}\;J\;{\cal{I}}{\it{m}}\left[\bar{J}(\chi)\right]
 }{\,\left(\;{\cal{R}}{\it{e}}\left[\bar{J}(\chi)\right]\;+\;{\cal{A}}_{\textstyle{_l}}\;J\;\right)^2\;+\;\left(\;{\cal{I}}{\it{m}}
 \left[\bar{J}(\chi)\right]\;\right)^2\,} ~~~.~~~~~
 \label{L39b}
 \end{eqnarray}
 Here $\,\chi\,$ is a shortened notation for the frequency $\,\chi_{\textstyle{_{lmpq}}}\,$, while the factors $\,{\cal{A}}_{\textstyle{_l}}\,$ are given by
 \begin{eqnarray}
 {\cal{A}}_{\textstyle{_l}}\,\equiv~\frac{\textstyle{(2\,{\it{l}}^{\,2}\,+\,4\,{\it{l}}\,+\,3)}}{\textstyle{{\it{l}}\,\mbox{g}\,
 \rho\,R}}~\mu
  ~=\;\frac{\textstyle{3\;(2\,{\it{l}}^{\,2}\,+\,4\,{\it{l}}\,+\,3)}}{\textstyle{4\;{\it{l}}\,\pi\,
 G\,\rho^2\,R^2}}~\mu~
  =~\frac{\textstyle{3\;(2\,{\it{l}}^{\,2}\,+\,4\,{\it{l}}\,+\,3)}}{\textstyle{4\;{\it{l}}\,
 \pi\,G\,\rho^2\,R^2~\,J
  }~}~\,~~~,\quad~~
 \label{L4}
 \end{eqnarray}
 $\rho\,$, g, and $\,R\,$ being the density, surface gravity, and radius of the body; and $\,G\,$ being the Newton gravitational constant. The unrelaxed elastic modulus and its inverse, the unrelaxed compliance, are denoted with $\,\mu\,$ and $\,J\,$, respectively. The complex compliance $\,\bar{J}(\chi)\,$ of the mantle is a Fourier image of the kernel $\,\dot{J}(t-t\,')\,$ of the integral equation
 \begin{eqnarray}
 2\,u_{\gamma\nu}(t)\,=\,\hat{J}(t)~\sigma_{\gamma\nu}\,=\,\int^{t}_{-\infty}\stackrel{\;\centerdot}{J}(t-t\,')~
 {\sigma}_{\gamma\nu}(t\,')\,d t\,'~~~~~
 \label{I12_4}
 \end{eqnarray}
 interconnecting the present-time deviatoric strain tensor $\,u_{\gamma\nu}(t)\,$ with the values assumed by the deviatoric stress $\,{\sigma}_{\gamma\nu}(t\,')\,$ over the time $\,t\,'\,\leq\,t\,$. The Fourier transform of (\ref{I12_4}) reads as:
 \begin{eqnarray}
 2\;\bar{u}_{\gamma\nu}(\chi)\,=\;\bar{J}(\chi)\;\bar{\sigma}_{\gamma\nu}(\chi)\;\;,
 \label{LLJJKK}
 \end{eqnarray}
 $\bar{u}_{\gamma\nu}(\chi)\,$ and $\,\bar{\sigma}_{\gamma\nu}(\chi)\,$ being the strain and stress in the frequency domain. The complex compliance $\,\bar{J}(\chi)\,$ contains contributions from elasticity, viscosity and inelastic processes (mainly, dislocation unjamming). Together, these three factors render the Andrade creep:
 \begin{subequations}
 \begin{eqnarray}
 {\bar{\mathit{J\,}}}(\chi)&=&J\,+\,\beta\,(i\chi)^{-\alpha}\;\Gamma\,(1+\alpha)\,-\,\frac{i}{\eta\chi}
  \label{112_1}\\
 \nonumber\\
 &=& J\,+\,\beta\,(i\chi)^{-\alpha}\;\Gamma\,(1+\alpha)\,-\,i\,J\,(\chi\,\tau_{_M})^{-1}
 ~~,
 \label{112_2}
 \end{eqnarray}
 $\Gamma\,$ denoting the Gamma function; $\,\eta\,$ being the mantle viscosity; $\,\tau_{_M}\equiv\eta/\mu=\eta J\,$ being the Maxwell time; $\,\alpha\,$ and $\,\beta\,$ being a dimensionless and dimensional Andrade parameters. The parameter $\,\beta\,$ has fractional dimensions, which makes it impractical; so it was suggested in Efroimsky (2012$\,$a, 2012$\,$b) to rewrite the compliance as
 \begin{eqnarray}
 {\bar{\mathit{J\,}}}(\chi)~=~J\,\left[\,1\,+\,(i\,\chi\,\tau_{_A})^{-\alpha}\;\Gamma\,(1+\alpha)~-~i~(\chi\,\tau_{_M})^{-1}\right]\;\;\;,
 \label{112_3}
 \end{eqnarray}
 \label{LL44}
 \end{subequations}
 with the parameter $\,\tau_{_A}\,$ defined through
 \begin{eqnarray}
 \beta\,=\,J~\tau_{_A}^{-\alpha}~~.
 \label{beta}
 \end{eqnarray}
 In {{Ibid.}}, $\,\tau_{_A}\,$ was christened $ \, ${\it{the Andrade time}}$ \, $.

 Below some threshold frequency (Karato and Spetzler 1990, Eqn. 17), dislocation unjamming becomes less efficient, and the rheology of the mantle becomes purely viscoelastic. This is why $ \, ${\it{at low frequencies}}$ \, $ it is legitimate to treat the mantle as the Maxwell body. Mathematically, this is expressed through the Andrade time rapidly growing as the frequency goes beneath the said threshold; so at lower frequencies the complex compliance becomes simply
 \begin{eqnarray}
 ^{\textstyle{^{(Maxwell)}}}{\bar{\mathit{J\,}}}(\chi)~=~J\,-\,\frac{i}{\eta\chi}~=~J\,\left[\,1~-~i~(\chi\,\tau_{_M})^{-1}\right]\;\;\;.
 \label{maxwell}
 \end{eqnarray}
 Insertion of this formula into the expression (\ref{L39b}) yields:
 \begin{eqnarray}
 ^{\textstyle{^{(Maxwell)}}}{k}_{\textstyle{_l}}(\chi)\;\sin\epsilon_{\textstyle{_l}}(\chi)\;=\;\frac{3}{2\,({l}\,-\,1)}\;\,\frac{{\cal{A}}_{\textstyle{_l}}\;(\chi\,\tau_{_M})^{-1}}{
 \,\left(\;1\;+\;{\cal{A}}_{\textstyle{_l}}\;\right)^2\;+\;(\chi\,\tau_{_M})^{-2}\,} ~~~.~~~~~
 \label{kosa}
 \end{eqnarray}
 In Section \ref{case3}, we use this formula to model dissipation in the planet Kepler-10$ \, $b.

 \end{document}